# On the compressibility of the transition-metal carbides and nitrides alloys $Zr_xNb_{1-x}C$ and $Zr_xNb_{1-x}N$

Vassiliki Katsika-Tsigourakou[a]

*Section of Solid State Physics, Department of Physics, National and Kapodistrian University of Athens, Panepistimiopolis, 157 84 Zografos, Greece*

**Abstract**

The 4d-transition-metals carbides (ZrC, NbC) and nitrides (ZrN, NbN) in the rocksalt structure, as well as their ternary alloys, have been recently studied by means of a first-principles full potential linearized augmented plane waves method within the local density approximation. These materials are important because of their interesting mechanical and physical properties, which make them suitable for many technological applications. Here, by using a simple theoretical model, we estimate the bulk moduli of their ternary alloys $Zr_xNb_{1-x}C$ and $Zr_xNb_{1-x}N$ in terms of the bulk moduli of the end members alone. The results are comparable to those deduced from the first-principles calculations.

--------------------------------------------------

[a]*E-mail address*: vkatsik@phys.uoa.gr

# 1. Introduction

A first principles study of 4d-transition-metal carbides (ZrC, NbC) and nitrides (ZrN, NbN) and their ternary alloys $Zr_xNb_{1-x}C$ and $Zr_xNb_{1-x}N$ (x=0.00, 0.25, 0.50, 0.75, 1.00) recently appeared [1]. The intensified interest in these transition metals (TMs) carbides and nitrides stems from the fact that they display a number of unique properties, including extremely high melting temperature and hardness, as well as high thermal and electrical conductivity and chemical stability [2]. The combination of such properties make these materials potential candidates for a variety of high temperature structural applications that gave rise to a number of theoretical studies [3-11] and experimental investigation [11-20].

Alloying in TM carbides and nitrides has not been much studied experimentally and the theoretical calculation of the chemical bonding for these systems is, therefore, of considerable interest. To learn more about the nature of these materials, Ref. [1] describes the mechanical and electronic behavior of a series of binary TM carbides and nitrides and their ternary alloys. Zaoui et al. [1] used an accurate full-potential density- functional method in the NaCl structure (in which most of these carbides and nitrides crystallize [21]), by employing the scalar relativistic full-potential linearized augmented plane wave method (FPLAPW) and the Perdew-Wang local density approximation [22]. They calculated all the equilibrium structural parameters and obtained, for the first time, values for all elastic constants of these materials in their equilibrium rocksalt phase. An inspection of these results (see Table 1 of Ref. [1]), shows that the values of bulk moduli of selected TM carbides and nitrides differ from those of the corresponding ternary alloys $Zr_xNb_{1-x}C$ and $Zr_xNb_{1-x}N$, (see also Fig. 1b of Ref. [1]) Therefore, the question arises whether one can determine this value of a mixed system solely in terms of the elastic data of the end members. This paper aims to answer this question. We employ here a simple model (described below in Section II), that has been also recently [23] used for the calculation of the compressibilities of multiphased mixed crystals grown by the melt method using the miscible alkali halides, i.e., NaBr and KCl, and measured in a detailed experimental study by Padma and Mahadevan [24,25].

## 2. The model

In Ref. [23], a model has been presented that explains how the properties of a mixed system $A_xB_{y-x}C_{1-y}$ can be determined in terms of the properties of the three end members A, B and C. Here, we recapitulate this model but for the case of a mixed system $A_xB_{1-x}$ which is of interest for the purpose of the present study. Following chapter 12 of Ref. [26] (based on Refs. [27,28]), we will explain how the compressibility of the mixed system $A_xB_{1-x}$ can be determined in terms of the compressibilities of the two (pure) end members $A$ and $B$. Let us call these two end members $A$ and $B$ as pure components (1) and (2), respectively and label $\upsilon_1$ the volume per "molecule" of the pure component (1) (assumed to be the major component in the aforementioned mixed system $A_xB_{1-x}$), $\upsilon_2$ the volume per "molecule" of the pure component (2). Furthermore, let $V_1$ and $V_2$ denote the corresponding molar volumes, i.e. $V_1 = N\upsilon_1$ and $V_2 = N\upsilon_2$ (where $N$ stands for Avogadro's number) and assume $\upsilon_1 < \upsilon_2$. Defining a "defect volume" $\upsilon_{2,1}^d$ as the increase of the volume $V_1$ if one "molecule" of type (1) is replaced by one "molecule" of type (2), it is evident that the addition of one "molecule" of type (2) to a crystal containing $N$ "molecules" of type (1) will increase its volume by $\upsilon_{2,1}^d + \upsilon_1$. Assuming that $\upsilon_{2,1}^d$ is independent of composition, the volume $V_{N+n}$ of a crystal containing $N$ "molecules" of type (1) and $n$ "molecules" of type (2) can be written as:

$$V_{N+n} = N\upsilon_1 + n(\upsilon_{2,1}^d + \upsilon_1), \quad \text{or} \quad V_{N+n} = [1 + (n/N)]V_1 + n\upsilon_{2,1}^d \qquad (1)$$

The compressibility $\kappa$ of the mixed crystal can be found by differentiating Eq.(1) with respect to pressure which gives:

$$\kappa V_{N+n} = [1 + (n/N)]\kappa_1 V_1 + n\kappa_{2,1}^d \upsilon_{2,1}^d \qquad (2)$$

where $\kappa_{2,1}^d$ denotes the compressibility of the volume $\upsilon_{2,1}^d$, i.e., $\kappa_{2,1}^d \equiv -(1/\upsilon_{2,1}^d) \times (d\upsilon_{2,1}^d/dP)_T$. Within the approximation of the hard-spheres model, the "defect–volume" $\upsilon_{2,1}^d$ can be estimated from:

$$\upsilon_{2,1}^d = (V_2 - V_1)/N \quad \text{or} \quad \upsilon_{2,1}^d = \upsilon_2 - \upsilon_1 \qquad (3)$$

Thus, since $V_{N+n}$ can be determined from Eq.(1) [upon considering Eq.(3)], the compressibility $\kappa$ can be found from Eq.(2) if a procedure for the estimation of the compressibility $\kappa_{2,1}^d$ of the "defect-volume" $\upsilon_{2,1}^d$ is employed. In this direction, we adopt a thermodynamical model for the formation and migration of the defects in solids described below, which has been of value in various categories of solids including metals, ionic crystals, rare gas solids, etc [29-33], as well as in high $T_c$ superconductors [34], and in complex ionic materials under uniaxial stress [35] that emit electric signals before fracture, similar to signals observed [36,37] before the occurrence of major earthquakes. According to the latter model, the defect Gibbs energy $g^i$ is interconnected with the bulk properties of the solid through the relation $g^i = c^i B\Omega$ (usually called $cB\Omega$ model) where $B$ stands for the bulk modulus ($=1/\kappa$), $\Omega$ the mean volume per atom and $c^i$ is dimensionless quantity. (The superscript $i$ refers to the defect process under consideration, e.g. defect formation, defect migration and self-diffusion activation). By differentiating this relation in respect to pressure $P$, we find that defect volume $\upsilon^i$ [$=(dg^i/dP)_T$] the compressibility $\kappa^{d,i}$ which, defined as $\kappa^{d,i} [\equiv -(d\ell n\upsilon^i/dP)_T]$, is given by [30]:

$$\kappa^{d,i} = (1/B) - (d^2B/dP^2)/[(dB/dP)_T - 1] \tag{4}$$

We now assume that the validity of Eq. (4) holds also for the compressibility $\kappa_{2,1}^d$ involved in Eq. (2), i.e.,

$$\kappa_{2,1}^d = \kappa_1 - (d^2B_1/dP^2)/[(dB_1/dP)_T - 1] \tag{5}$$

where the subscript "1" in the quantities at the right side denotes that they refer to the pure component (1). The quantities $dB_1/dP$ and $d^2B_1/dP^2$, when they are not experimentally accessible, can be estimated from the modified Born model according to [27,26]:

$$dB_1/dP = (n^B + 7)/3 \text{ and } B_1(d^2B_1/dP^2) = -(4/9)(n^B + 3) \tag{6}$$

where $n^B$ is the usual Born exponent. This is the procedure that has been successfully applied in Ref. [23] for the multiphased mixed alkali crystals. Let us call it, for the sake of convenience, Procedure 1.

An alternative procedure (hereafter called Procedure 2) for the estimation of the compressibility of a mixed system from the compressibility of the end members, is as follows [26-28]:

Equation (1) gave the volume resulting from the addition of $n$ molecules to a pure body consisting of $N$ molecules. It can be set in a different form by considering the volume $V$ in which $n$ molecules have replaced $n$ molecules of the pure body. The molar fraction $x$ is connected to $n/N$ by: $n/N = x/(1-x)$. Then using Eqs. (1) and (3) one gets:

$$V = (1-x)V_1 + xV_2 \tag{7}$$

Differentiating Eq. (7) with respect to pressure we get:

$$\kappa V = (1-x)\kappa_1 V_1 + x\kappa_2 V_2 \qquad (8)$$

where $\kappa$, $\kappa_1$ and $\kappa_2$ are the compressibility for the mixed system and the pure components (1) and (2) respectively. Inserting the value of $V$ from Eq. (7) we find the bulk modulus $B$ ($\equiv 1/\kappa$)

$$B = B_1 \frac{1 + x[(V_2/V_1) - 1]}{1 + x[(B_1 V_2 / B_2 V_1) - 1]} \qquad (9)$$

This equation permits the direct evaluation of $B$ at any desired composition and temperature in terms of the elastic data of the two end members.

Both the aforementioned procedures assume that no additional aliovalent impurities are present in the mixed system that may influence the dielectric and electrical properties [38].

## 3. Results of the model

Here, we use the calculated values for the lattice constants ($a$), bulk moduli ($B$) and their first pressure derivatives ($dB/dP$) of ZrC, ZrN, NbC and NbN, given in Table 1 of Ref. [1]. Concerning the second pressure derivatives ($d^2B/dP^2$), which are not given in Ref. [1], they are estimated here from the relations (6) through the Born coefficients $n^B$. Below we report our results calculated on the basis of the two procedures described in Section II. All the results are presented in Table 1. They are also plotted in Figs. 1(a) for $Zr_xNb_{1-x}C$ and 1(b) for $Zr_xNb_{1-x}N$, respectively.

*3.1. Application of the model to $Zr_xNb_{1-x}C$ (x= 0.25, 0.50, 0.75)*

Let us start with the application of the Procedure 1. For x=0.25, the end member (pure) crystal with the higher composition is NbC (component (1)). We use the values of $B(\equiv 1/\kappa)$=338.158 GPa, and $dB/dP$=4.192 for NbC, which are given in the Table 1 of A. Zaoui et al. [1] and using Eq. (6), we find $d^2B/dP^2$=-0.0113 GPa$^{-1}$. By inserting these values into Eq. (5) we find

$$\kappa_{2,1}^d = 6.497 \times 10^{-3} \text{ GPa}^{-1}$$

Furthermore, by considering the $\upsilon_1$ and $\upsilon_2$ values of NbC and ZrC respectively, we find $\upsilon_{2,1}^d = 3.262 \times 10^{-24} cm^3$ from Eq. (3) and $V_{N+n} = 30.145 \times 10^{-24} cm^3$ from Eq. (1). By inserting the aforementioned values into Eq. (2) we find $\kappa = 3.085 \times 10^{-3}$ GPa$^{-1}$ and therefore $B = 324.149$ GPa.

For x=0.50 and considering that NbC as component (1), we find $V_{N+n} = 46.848 \times 10^{-24} cm^3$ and $B = 312.152$ GPa by following the same procedure. In the case where ZrC is component (1), and from the values $B(\equiv 1/\kappa)$=247.508 GPa and $dB/dP$=4.029, which are given in the Table 1 of A. Zaoui et al. [1], we find $d^2B/dP^2$=-0.0145 GPa$^{-1}$ and $\kappa_{2,1}^d = 8.827 \times 10^{-3}$ GPa$^{-1}$. Furthermore, by considering the $\upsilon_1$ and $\upsilon_2$ values of ZrC and NbC respectively, we find $\upsilon_{2,1}^d = -3.262 \times 10^{-24} cm^3$ from Eq. (3) and finally $B = 269.784$ GPa.

For x=0.75, the end member (pure) crystal with the higher composition is ZrC. Following the above procedure and the corresponding values for ZrC, as component (1), we find $V_{N+n} = 32.319 \times 10^{-24} cm^3$ and finally $B = 257.651$ GPa.

We now turn to the application of the Procedure 2. By applying Eq. (9), we find the following values for the bulk moduli: $B$=307.389, 282.767 and 263.380 GPa for x=0.25, 0.50 and 0.75 respectively. The results are the same when either ZrC or NbC is considered as component (1).

*3.2. Application of the model to $Zr_xNb_{1-x}N$ (x= 0.25, 0.50, 0.75)*

We start again with the application of the Procedure 1. For x=0.25, the end member (pure) crystal with the higher composition is NbN (component (1)). We use the values of $B(\equiv 1/\kappa)$=355.262 GPa and $dB/dP$=4.705 for NbN, which are given in the Table 1 of A. Zaoui et al. [1], and using Eqs (6), we find $d^2B/dP^2$=-0.0127 GPa$^{-1}$. By inserting these values into Eq. (5) we find

$$\kappa_{2,1}^d = 6.243 \times 10^{-3} \text{ GPa}^{-1}$$

Furthermore, by considering the $\upsilon_1$ and $\upsilon_2$ values of NbN and ZrN respectively, we find $\upsilon_{2,1}^d = 2.447 \times 10^{-24} cm^3$ from Eq. (3) and $V_{N+n} = 28.519 \times 10^{-24} cm^3$ from Eq. (1). By inserting the aforementioned values into Eq. (2) we find $\kappa = 2.913 \times 10^{-3}$ GPa$^{-1}$ and therefore $B = 343.288$ GPa.

For x=0.50 and considering that NbN is the component (1), we find $V_{N+n} = 44.002 \times 10^{-24} cm^3$ and $B = 332.706$ GPa. In the case where ZrN is component (1), and from the values $B(\equiv 1/\kappa)$=286.852 GPa and $dB/dP$=4.111, which are given in the Table 1 of A. Zaoui et al. [1], we find $d^2B/dP^2$=-0.0129 GPa$^{-1}$ and $\kappa_{2,1}^d = 7.633 \times 10^{-3}$ GPa$^{-1}$. Furthermore, by considering the $\upsilon_1$ and $\upsilon_2$ values of ZrN and NbN respectively, we find $\upsilon_{2,1}^d = -2.447 \times 10^{-24} cm^3$ from Eq. (3) and finally $B = 307.188$ GPa.

For x=0.75, the end member (pure) crystal with the higher composition is ZrN. Following the above procedure and the corresponding values for ZrN as component (1), we find $V_{N+n} = 30.150 \times 10^{-24} cm^3$ and finally $B = 296.398$ GPa.

We now turn to the application of Procedure 2. By applying Eq. (9), we find the following results: $B$ =333.660, 315.540 and 300.124 GPa for x=0.25, 0.50 and 0.75 respectively. The results are the same when either ZrN or NbN is considered as component (1).

## 4. Conclusion

Using two different procedures that are based on a simple thermodynamical model, we estimated the bulk moduli of ternary alloys $Zr_xNb_{1-x}C$ and $Zr_xNb_{1-x}N$ in terms of the bulk moduli of their end members. Both procedures lead to bulk moduli values which are comparable with those obtained from detailed first-principles calculations. In particular, the differences are small, i.e., about (1.5-8)% for $Zr_xNb_{1-x}C$ and (0.7-6)% for $Zr_xNb_{1-x}N$, for the procedure 1, while for the procedure 2 the differences are (3.6-3.8)% for $Zr_xNb_{1-x}C$ and (0.5-0.8)% for $Zr_xNb_{1-x}N$.

**Table 1**. The values of the bulk modulus calculated by both procedures compared to those obtained from the first-principles study in Ref. [1].

| composition | x | $B^a$ (GPa) | $B^b$ (GPa) | $B^c$ (GPa) |
|---|---|---|---|---|
| $Zr_xNb_{1-x}C$ | 0.00 | 338.158 | | |
| | 0.25 | 318.911 | 324.149 | 307.389 |
| | 0.50 | 294.007 | 312.152 | 282.767 |
| | | | 269.784 | |
| | 0.75 | 273.701 | 257.651 | 263.380 |
| | 1.00 | 247.508 | | |
| $Zr_xNb_{1-x}N$ | 0.00 | 355.262 | | |
| | 0.25 | 331.586 | 343.288 | 333.660 |
| | 0.50 | 313.020 | 332.706 | 315.540 |
| | | | 307.188 | |
| | 0.75 | 298.486 | 296.398 | 300.124 |
| | 1.00 | 286.852 | | |

[a]Literature values, which are given from Table 1 and Fig. 1b (Ref. [1])

[b]Calculated from Eq. (2) by inserting $\kappa^d$ from Eq. (5), i.e., Procedure 1.

[c]Calculated from Eq. (9), i.e., Procedure 2.

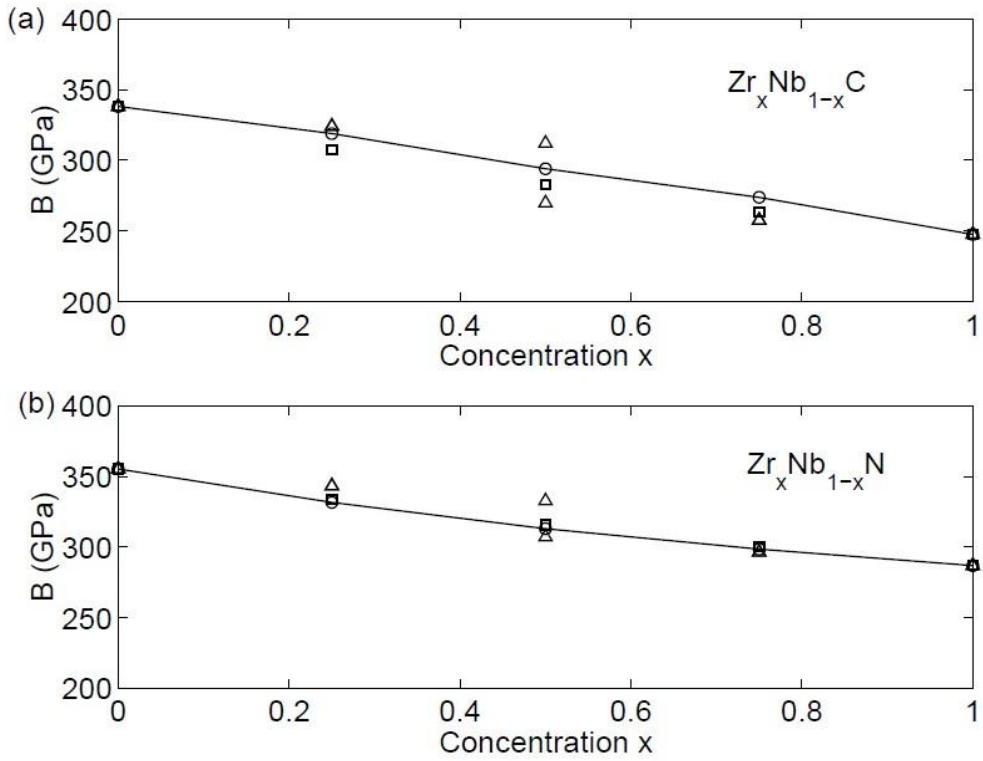

**Figure 1**. Variation of the bulk modulus (B) as a function of the concentration x, for: (a) $Zr_xNb_{1-x}C$ and (b) $Zr_xNb_{1-x}N$. Circles: calculated from Ref. [1]; Triangles and squares correspond to the values calculated using procedures 1 and 2, respectively (see the text).